\documentclass[acmsmall]{acmart}
\usepackage{csquotes}
\usepackage{xcolor}
\usepackage{ulem}

\AtBeginDocument{%
  \providecommand\BibTeX{{%
    \normalfont B\kern-0.5em{\scshape i\kern-0.25em b}\kern-0.8em\TeX}}}

\setcopyright{acmcopyright}
\copyrightyear{2023}
\acmYear{2023}
\acmDOI{XXXXXXX.XXXXXXX}

\acmJournal{JACM}
\acmVolume{37}
\acmNumber{4}
\acmArticle{111}
\acmMonth{8}

\begin{document}

\title{Towards Intersectional Moderation: An Alternative Model of Moderation Built on Care and Power}

\author{Sarah Gilbert}
\email{sarah.gilbert@cornell.edu}
\orcid{0000-0003-2718-4121}
\affiliation{
\institution{Cornell University}
\city{Ithaca}
\state{NY}
\country{USA}
}

\renewcommand{\shortauthors}{Gilbert}

\begin{abstract}
Shortcomings of current models of moderation have driven policy makers, scholars, and technologists to speculate about alternative models of content moderation. While alternative models provide hope for the future of online spaces, they can fail without proper scaffolding. Community moderators are routinely confronted with similar issues and have therefore found creative ways to navigate these challenges. Learning more about the decisions these moderators make, the challenges they face, and where they are successful can provide valuable insight into how to ensure alternative moderation models are successful.   

In this study, I perform a collaborative ethnography with moderators of r/AskHistorians, a community that uses an alternative moderation model, highlighting the importance of accounting for power in moderation. Drawing from Black feminist theory, I call this  ``intersectional moderation.'' I focus on three controversies emblematic of r/AskHistorians' alternative model of moderation: a disagreement over a moderation decision; a collaboration to fight racism on Reddit; and a period of intense turmoil and its impact on policy. Through this evidence I show how volunteer moderators navigated multiple layers of power through care work. To ensure the successful implementation of intersectional moderation, I argue that designers should support decision-making processes and policy makers should account for the impact of the sociotechnical systems in which moderators work. 
\end{abstract}



\maketitle

\section{Introduction}
Legend has it that r/AskHistorians, a Reddit community well known for its strict moderation practices, wasn’t always that way. Founded in 2011 by Reddit user u/artrw, r/AskHistorians was created to provide a place for people to ask questions about history and get informed answers. u/artrw was libertarian and therefore believed a hands-off approach to moderation was best. The small yet active community developed norms encouraging in-depth and comprehensive answers to users’ questions. However, as the community’s high-quality discussions began to attract more users, it became clear that social norms alone would not be enough to maintain the quality of discussions its original members enjoyed. As the community grew, they began to witness hate-based myths and conspiracies such as Holocaust denial \cite{breit2018how}, the Lost Cause, and American Indian Genocide denial upvoted and propagated. Reddit had earned an unsavory reputation at the time. Free speech was a core value held by Reddit’s administrators and its users, and moderation techniques such as removing content violated that ethos \cite{chen2012reddit}. The result was what Massanari \cite{massanari2017gamergate} describes as a ``toxic technoculture.'' On Reddit, a platform where free speech was the reigning paradigm, r/AskHistorians developed an alternative moderation model in an attempt to counter the injustices they witnessed on Reddit while promoting public history \cite{raeburn2022out, gilbert2020run}.

Currently, online moderation lies in the hands of platforms and users that use a variety of content moderation models \cite{klonick2017new, roberts2019behind, gillespie2018custodians}. But these come at a cost. For example, although top-down, centralized approaches to moderation, such as banning users, can reduce hate speech \cite{jhaver2021evaluating}, bans may also disproportionately impact people who have been historically marginalized, \cite{gray2021we, thach2022visible}, and civility moderation algorithms have been found to perpetuate racism and misogynoir \cite{marshall2021algorithmic, buolamwini2018gender, hamilton2019biased, sandvig2016automation}. While bottom-up, decentralized approaches, such as voting, can involve more people in moderation decisions \cite{lampe2004slash, lampe2014crowdsourcing}, they can also silence people who are marginalized and promote dominant viewpoints \cite{massanari2017gamergate}. In response, scholars have called for alternative content moderation models that center harm reduction \cite{phillips2021you, schoenebeck2021drawing, schoenebeck2020reimagining, habib2022proactive}. However, we know little about the processes involved in moderation that reduce harm, which places well-intentioned solutions at risk of failure. I argue that if moderation is going to avoid reproducing harm, it must account for power. Drawing from Black Feminist theory \cite{collins1990black, crenshaw2018demarginalizing}, I refer to this as ``intersectional moderation.'' 

In this paper I use collaborative ethnography to analyze three controversies that are emblematic of challenges practicing alternative moderation models and accounting for power. The first controversy describes a disagreement over a moderation decision; the second describes two concurrent instances during which the moderation team collaborated to fight racism on Reddit; and the third describes an instance of intense turmoil and its impact on policy. These controversies highlight how r/AskHistorians moderators engage in care work, defined broadly as labor undertaken out of affection or a sense of responsibility for others \cite{folbre1995holding}, to navigate layers of power outlined by Black Feminist scholar, Patricia Hill Collins \cite{collins1990black}. Using these findings, I highlight platform policy and design gaps that make it difficult for moderators to enact intersectional moderation and provide suggestions for implementing it. 

\section {Background}
\subsection{Existing models of content moderation}
To understand the call for alternate moderation models it is important to know how moderation is currently operationalized. I have organized existing models according to their hierarchical structure to focus on the source of power; is moderation enacted from the top down (conducted by the platform or another central authority) or the bottom up (conducted by regular users)?  I then describe each of these models' existing tactics and approaches, before advancing an intersectional model of content moderation, bolstered by theories of intersectionality \cite{crenshaw2018demarginalizing, collins1990black}.  

\subsubsection{Top-down moderation models}
Top-down approaches to governance are directed and enacted by a central authority. They are conducted either at the platform level, or by users given moderator or administrator status. Top-down approaches include individual-level sanctions (such as removing content posted by users and banning users), community-level sanctions (in which entire communities are isolated or banned), algorithmic downranking, and some forms of fact-checking and labeling controversial contentment. This review focuses on individual-level and community-level sanctions as these are types of moderation commonly used by community moderators across platforms and/or may impact Reddit community moderators specifically. 

Prior research on individual-level sanctions such as \textit{content removal} has shown mixed impacts on improving online community health. For example, while removing content can have an immediate impact on harm reduction, it may not benefit communities in the long term \cite{srinivasan2019content}. One reason may be because when users are unaware of why content was removed, the community never develops beneficial social norms. While normative information can show users how to act in line with a group \cite{tankard2016norm}, they are rarely aware that their content was removed \cite{jhaver2019did} and rely on folk theorization to develop explanations for moderation practices \cite{myers2018censored}. Indeed, making users aware of community norms by ``stickying'' a comment to the top of a thread reduces rule-violating comments \cite{matias2019preventing} and users given explanations for content removal are less likely to have a post removed in the future \cite{jhaver2019does}. Explicit rules increase perceptions of fairness \cite{tyler2021social} and provide learning opportunities \cite{thach2022visible}. However, providing explanations for removals may place community content moderators at risk. Making moderation transparent to users places pressure on moderators, making them vulnerable to harassment and abuse if users respond negatively to a removal notice \cite{gilbert2020run, dosono2019moderation}.

\textit{Banning users}—also referred to as ``deplatforming''—is another form of top-down moderation. Bans can be either temporary or permanent removals from a platform. Temporary bans may serve several purposes, such as preventing users from repeatedly violating rules, giving them a punitive ``time-out,'' or deterring future rule-breaking behaviour. However, using temporary bans as a warning system may not prevent users from engaging in future rule-breaking behavior and may risk alienating users from contributing content in the future \cite{chang2019trajectories}. Permanently banning users prevents people from participating on a particular platform or within a particular community. While bans are a particularly serious sanction, the ease with which users of many platforms can create new accounts can reduce their efficacy. For example, bans are easily evaded on platforms where participation is pseudonymous, prompting the development of systems to detect ban evasion (e.g., \cite{niverthi2022characterizing}). However, bans are more impactful when users participate with their real name (e.g., for publicity, commerce, and social connection) and when users occupy influential positions in social networks that allow them to broadcast information to wide audiences. Jhaver et al. \cite{jhaver2021evaluating} analyzed the impact of banning three influencers known for promoting offensive content, conspiracy theories, and engaging in targeted harassment. They found that after each user was banned by Twitter, offensive speech posted by their followers decreased. 

While bans may limit harmful content online, concerns remain about their disproportionate impact on historically marginalized people. Gray and Stein \cite{gray2021we} refer to these as ``carceral logics,'' whereby Black women are negatively impacted by ostensibly protective moderation measures. Instead, surveillance measures disproportionately target Black women, who, when they speak out against racism and sexism, are placed in ``Facebook jail.'' Similarly, Marshall \cite{marshall2021algorithmic} notes the white, colonist roots of content moderation, in which policies and the algorithms that enforce them are standardized and applied without nuance, such as ``civility'' algorithms that remove content calling out racism, while allowing the racist content to remain, a phenomenon she refers to as algorithmic misogynoir. Further, bans have disproportionate impacts on people who make their living online, particularly those who are already marginalized, such as those who engage in sex (or related) work \cite{are2023emotional}. The prioritization of powerful entities over affected users results in what Are \cite{are2022autoethnography} describes as ``automated powerlessness.'' 

In addition to sanctioning individual users, platforms may also sanction entire communities. For example, they might ban Facebook groups or subreddits, or isolate them in quarantines. As with deplatforming influential individuals, \textit{banning communities} has also been found to reduce hate speech \cite{chandrasekharan2017you}. However, banning communities on one platform does not keep communities from reestablishing themselves elsewhere. While new communities may see decreased rates of posting, they may also encounter increasing toxic speech and radicalization \cite{ribeiro2021platform}. Although evidence suggests that community-level sanctions help decrease toxicity on the original platform, they are also relatively rare. On Reddit, they are typically conducted in response to negative media attention \cite{habib2021reddit}. Community-level sanctions also highlight a power imbalance between administrators and users. While the primary purpose of subreddits is communicative, many also use Reddit’s features (e.g., FAQs and wikis) to archive information \cite{dosono2020decolonizing, squirrell2019platform}. A wrongfully sanctioned community risks losing years of discussions and resources \cite{dosono2020decolonizing}. 

\textit{Quarantines} are a middle ground between banning communities and allowing them to operate freely. Reddit introduced quarantines to limit the accidental discovery of controversial communities by placing content within the appropriate context \cite{reddit2021quarantines}. When a community is placed under quarantine, users must opt in to seeing the content. Quaratined communities generate no ad revenue, don't appear in non-subscription-based feeds such as r/all or r/popular, aren't included in searches, and aren't recommended by the platform \cite{reddit2021quarantines}. As with impacts of community bans, placing communities under quarantine has been found to successfully reduce the number of newcomers who may find content objectionable. However, they may not reduce levels of toxicity \cite{chandrasekharan2022quarantined, shen2022tale} and can have homogenizing impact on discussions within quarantined communities \cite{shen2022tale}. 

\subsubsection{Bottom-up moderation models}
Unlike community and commercial moderation, in which small groups of moderators or administrators engage in sanctioning, bottom-up moderation is conducted by users with no special privileges. The two most common forms of bottom-up moderation models are distributed moderation and individual-level controls such as blocks and filters. 

In \textit{distributed moderation,} users make judgements about the quality of content. For example, on Slashdot, users could choose from a list of descriptors such as ``offtopic,'' ``troll,'' or ``insightful'' to apply to content. These descriptors would impact the display order and could be used to filter content. Similarly, Reddit uses a voting system in which users vote on posts or comments and the total score (or ``karma'') determines how content is displayed. Highly upvoted content is promoted to the top of threads, subreddit pages, personal feeds, and site-wide feeds (such as r/popular or r/all), while downvoted content is hidden. Distributed moderation systems are typically easy to use, and therefore widely adopted by users. They also provide information about what content is accepted by and interesting to a community \cite{mills2018pop, lampe2004slash}. Distributed moderation may be thought of as more ``democratic,'' since a greater number of users participate in moderation. However, while these systems have advantages, they can also lead to biased content management \cite{massanari2017gamergate}. For example, in distributed moderation comments with lower scores may receive slower moderation, and incorrect moderation may not be reversed \cite{lampe2004slash}. Distributed moderation can also propagate misinformation when voters are not subject matter experts \cite{gilbert2020run}, reinforce echo chambers \cite{mills2018pop}, and push marginalized users further to the margins \cite{das2021jol}. 

In both top-down and distributed moderation, decisions about what content is allowed or banned, promoted or demoted, and who can participate affect all users. However, with \textit{individual-level controls}, such as  block lists and filters, individuals moderate the content they see and the people they interact with. However, these decisions only impact their experience of a platform. Blocking users is often the first line of defense for users facing online harassment. However, blocking—particularly in cases of networked harassment \cite{marwick2021morally}—can be time consuming and exhausting \cite{geiger2016bot}. Similarly, developing lists of filters is a challenging, ongoing task that demands user labor \cite{jhaver2022designing}. 

\subsection{Beyond bans, blocks, and removals: Alternative models of content moderation}

Many of models of moderation come with tradeoffs. In some cases, moderation approaches replicate societal harms enacted on marginalized people \cite{gray2021we, marshall2021algorithmic, thach2022visible, haimson2021disproportionate, are2020instagram, klassen2021more}. Further, the problems that moderation purports to solve still exist and the lack of transparency in current moderation models and the ability of platforms to effectively moderate have led to decreased user trust in moderation \cite{suzor2019we,wohn2019volunteer}. Issues with current moderation models and harms perpetuated through internet use have led scholars and technologists to rethink current moderation models and propose alternatives. This section discusses three alternative moderation models: proactive, justice-based, and networked ethics. 

The aim of \textit{proactive} moderation models is to identify and curtail content that is likely to be harmful before it can cause harm. Proactive models can be difficult to implement because they require prediction and human labor, and risk backlash. However, recent work by Habib et al. \cite{habib2022proactive} suggests that, within the context of Reddit, community-level proactive moderation empirically identified problematic communities earlier and with less human effort. To do so, the authors analyzed which users were once active in banned communities, the percentage of toxic comments, negative mentions in other communities, percentage of comments removed, and negative press coverage. Proactive moderation has also yielded promising results at the content level as well. For example, Ribeiro, Cheng and West \cite{ribeiro2022post} found that Facebook Groups in which moderators approved posts had fewer, but higher-quality posts. Further, community members were more active and reported posts less. Finally, moderators also engage in norm setting through participation in their communities, such as building relationships \cite{lo2018all} and providing content \cite{gilbert2020run}. However, as top-down moderation, without careful implementation, proactive models risk replicating harms found in other top-down models. 

\textit{Justice models} of moderation are concerned with repairing harms such as psychological distress, physical violence, oppression, marginalization, and threats to free expression. As Salehi \cite{salehi2020no} notes, framing online harms as a content moderation problem ``assumes that the problem is individual pieces of content to be moderated—not people and their relationships.'' In contrast, justice models focus on people and emphasis is placed on accountability and reparation to victims of online harms \cite{schoenebeck2020reimagining, salehi2020no}. For example, rather than top-down paternalistic models that can replicate the carceral logics described by Gray and Stein \cite{gray2021we}, justice-based models foster education, rehabilitation, and forgiveness. Sanctions would be proportionate to the violation, and decisions would account for context, focusing on behaviour rather than content. Justice models also foreground those who have been harmed to locate appropriate reparations \cite{schoenebeck2021drawing, schoenebeck2021youth}. Finally, justice-based models move beyond ``neutral'' decision-making frameworks to frameworks that support communities making moderation decisions for themselves \cite{schoenebeck2020reimagining, salehi2020no}.

Another alternative moderation model, \textit{network ethics}, suggests that we reconceptualize how moderation is enacted by questioning how a fundamental value of online communication—freedom of speech— should be operationalized. Using metaphors of ``ecology'' to describe the online landscape, Phillips and Milner \cite{phillips2021you} propose addressing the online pollution crisis through a collective response. They advocate for moving away from the internet's original libertarian approaches to freedom (negative freedoms) and instead think in terms of positive freedoms. Negative freedoms focus on the individual, assuring them freedom \textit{from} restriction, whereas positive freedoms focus on the community, assuring freedom \textit{for} the collective: ``It's the difference between assuring that an individual has the right to spew whatever poison they want without restraint, and asserting that those within the collective have the right not to be poisoned'' p.8. A disadvantage of negative freedoms is that they may have chilling effects \cite{niehoff2022free}—for example, if unrestrained bigotry crowds out participation among those who have been historically marginalized. However, a common critique of positive freedom is that it risks sliding into authoritarianism \cite{berlin2017two}, a concern in online contexts because authoritarian and paternalism can also have a chilling effect on participation \cite{matias2020automated} and can replicate systemic harms experienced by marginalized people \cite{gray2021we,thach2022visible, are2020instagram, haimson2021disproportionate, klassen2021more}. Therefore, positive freedom can only occur when inequality is addressed \cite{taylor2017s} and why networked ethics, like justice-based models, also involve broadening notions of harm and being accountable for harm \cite{phillips2021you}. 

\subsection{Towards an intersectional model of content moderation}

As described above, existing moderation paradigms fail to solve issues such as online harassment, abuse, and discrimination. To counter these shortcomings, alternatives have been proposed. I argue that in addition to, and/or in conjunction with these alternatives, moderation must also account for power. Existing models often re-enforce rather than subvert power relationships. For example, because it relies on a central authority, top down moderation can replicate paternalistic power structures while bottom up moderation can crowd out marginalized perspectives. 

Drawing from Black Feminist Theory and Kishonna Gray's Intersectional Tech framework \cite{gray2020intersectional}, \textit{intersectional moderation} would account for the impact of power across multiple levels of domination and arenas of resistance to oppression. In her groundbreaking work in which she coined the term ``intersectionality,'' Kimberlé Crenshaw \cite{crenshaw2018demarginalizing} described how anti-discrimination laws excluded Black women because they are multiply-burdened. Laws protecting race don't protect Black women because they aren't men, while laws protecting women don't protect Black women because they aren't white. Decisions in law, like decisions in moderation, rarely account for multiple axes of identities and experiences. Such universalist rules, guidelines, and applications can erase certain groups of people \cite{costanza2020design}, or make them highly visible targets \cite{devito2022how}. To explain how injustice affects people across multiple axes, Patricia Hill Collins \cite{collins1990black} describes oppression within a matrix of domination, in which each person derives varying amounts privilege and penalty from systems of oppression. Domination, particularly through racism, manifests across varying levels of power: personal (power upheld within relationships), community (power developed within cultural contexts and upheld by groups), and systemic (power upheld by social  institutions). While each of these levels of power are sites of domination, that they are also sites of resistance \cite{collins1990black, gray2020intersectional} An intersectional model of moderation would bring awareness to how moderation reinforces existing power structures at each level so that moderation policies and practices can subverted from reinforcing oppression to supporting resistance. 

r/AskHistorians has a mission: public history \cite{gilbert2020run, raeburn2022out}. To support this mission, r/AskHistorians uses an innovative model of moderation that includes elements of the alternative models described above. For example, by fostering education through question answering that closes gaps in empathy between Reddit's audience and people in the past \cite{gilbert2020run}, r/AskHistorians moderators use proactive models and justice models; through removals that hide hate-based disinformation and create space for people to write in-depth and comprehensive responses, r/AskHistorians centers positive freedom for the collective. The r/AskHistorians moderation community uses alternative moderation to promote public history and in doing so, negotiates power on a regular basis. However, the alternative model used by r/AskHistorians is only one way, one group, on one platform that attempts to account for and is challenged by power in moderation. r/AskHistorians' moderation model is not, and should not considered a definitive model of intersectional moderation. Rather, this study provides insight into how moderators are affected by power, challenges they experience negotiating it, and the role of platform design in these negotiations. Future work with communities should explore practices accounting for power through moderation across diverse contexts; how these practices can be supported through design and policy; if (or how) intersectional moderation might scale; and the impact of intersectional moderation on those who are marginalized and harmed by moderation.   

\section{Methods}
Prior work on moderation labor has shown that despite the high visibility of certain actions, such as interacting with community members, a great deal of moderation work is invisible \cite{dosono2019moderation, dosono2020decolonizing, gilbert2020run, li2022all, seering2019moderator}. While the rich results of studies such as these and others (e.g., \cite{matias2019civic, jiang2019moderation, kiene2016surviving, wright2022automated, squirrell2019platform, li2022measuring}) that use interviews, observation, and log analyses to provide quantitative and qualitative insights into community moderation labor, ethnographic methods can build on this work by providing insight into deliberative processes and procedures. Lo \cite{lo2018all}, for example, draws upon their experience as a Reddit moderator to advance a more holistic model of moderation that includes proactive community engagement. Rather than engage in observation alone, which could be considered extractive \cite{smith2021decolonizing} I opted to use collaborative ethnographic \cite{lassiter2005chicago} methods so that community members could be as involved in the research process as they wanted and so that I could give back through contributions to the community, such as providing skills (e.g., survey development) and labor (by contributing to the work involved in moderation). 

In January 2020 I approached r/AskHistorians moderators through the ``modmail`` feature of Reddit in which I proposed a collaborative ethnography to learn about moderation labor. In addition to requesting access to moderation spaces, I also requested to be added as a moderator. As with the addition of all new moderators, the r/AskHistorians moderation team held a discussion and voted, resulting in my addition as moderator and researcher. After learning that the community was amenable to participation in the project, I sought and was granted IRB approval.

\subsection{Data collection, analysis, and ethics}
Data collection was conducted across several online spaces: a private moderation subreddit (r/askhistoriansmods or ``modsub''), a private Slack workspace, and the subreddit (r/AskHistorians). While many discussions on r/AskHistorians are public, as a mod I could also see non-public content. This content included removed posts and comments, and discussions between moderators and users in the private messaging system, modmail. The private subreddit, r/askhistoriansmods, is used to discuss subreddit business and operating decisions, particularly when a record of discussions is required. Most discussions between moderators occur on a Slack workspace, which is divided into ``social'' and ``work'' channels. Social channels include \#funtime-channel (or ``funchan''), the default channel for general discussions; \#politicaldiscussion (or ``polchan''), for discussing politics; and \#writing\_help, for discussing writing and job-related challenges. Work channels include \#business-chat (or ``bizchan''), the main channel for discussing day-to-day moderation decisions; \#modschool for onboarding new mods; and \#modmail-tracker, which uses a bot to share links of all modmail messages to the channel and is used to coordinate responses. Former moderators who leave the team on good terms continue to engage in the Slack’s social spaces. While 79 people currently have access to the Slack, around 60 were active users over the course of this study. The mod team fluctuates in size, particularly because the team encourages taking a hiatus from moderating duties to mitigate burnout. As of writing there are currently 43 r/AskHistorians moderators; however, some are inactive.

I collected data from late January 2020 to late March 2021. I took screenshots and saved links where discussions were persistent (i.e., on Reddit). However, because Slack is ephemeral and thus could be considered more private, I opted not to take screenshots and instead took detailed notes. While writing this paper I requested access to Slack transcripts from particular date ranges so that I could corroborate details from my notes, fill gaps in my knowledge, and verify dates and times. In addition to describing the discussions, my notes also included my reflections via quick observations and longer reflective memos. On several occasions, I shared  memos with the moderation team, who helped confirm and refine emergent themes. Following grounded theory \cite{corbin2014basics, charmaz2006constructing}, through this process of memoing I identified themes involving different types of care work: care for each other, care for users, and care for the community, and the role of power in moderation. I then reviewed my notes to identify three controversies that were emblematic of these themes. After submitting the paper for review, I returned to the themes. Using axial coding \cite{corbin2014basics} and Collin's Matrix of Domination \cite{collins1990black}, I mapped themes of moderation care work to three levels of power: personal, community, and systemic.   

Collaborative ethnography includes participants throughout the research process \cite{lassiter2005chicago, fluehr2008collaborative}. Therefore, moderators were offered a voice in the paper if they wanted it and had a say in what was published. As Tuhiwai Smith \cite{smith2021decolonizing} notes, ``sharing knowledge is a long term commitment'' requiring reciprocity and feedback (p. 16). Accordingly, after drafting the Methods and Results sections, I shared a draft of the paper with the moderation team. Collaborative methods, including collaborative ethnography, should also center the voices of those who are impacted \cite{costanza2020design}. For example, Lassiter \cite{lassiter2005chicago} recommends including community members as authors. However, the bar for authorship can be high, potentially requiring more time than community members have to offer. To involve the r/AskHistorians community in the writing process while also accounting for members' availability and interest, I offered a compromise, inviting the moderation team to include their thoughts and reflections on each of the three controversies as footnotes, including my own reflections as examples. Six people provided annotations across the three controversies. In addition to being inclusive of the community with which I worked, these reflections also provide important information about care work and power.  While I recommend reading the annotations in the context of the controversies, readers may prefer to read them after each controversy. After the moderation team had a chance to read the paper but before comments were added, publishing it was put to a vote. Details on the vote are included in the Appendix. 
 
Because some moderators may be identifiable to people in the broader r/AskHistorians community, by default I used pseudonyms to protect their privacy to the greatest extent possible. I also chose anglicized names to prevent re-identification though race, ethnicity, or location clues, and obfuscated key details in some controversies. Prior to publication, I invited moderators to choose how they wished to be attributed in the paper and offered the option to vary attribution between controversies. If there was an incident involving them that they wanted obfuscated further, I honored that request. While I attempted to maintain an inclusive and democratic process, this was not fully possible. Several moderators who are included in the controversies left the moderation team prior to writing. In some cases, I was unsure of how to reach them and in others I didn't feel safe reaching out. While I considered not telling these stories, each of those moderators had agreed to add me as a moderator and allow me conduct research in the space, and I opted to follow ethics of care \cite{botes2000comparison} by prioritizing the collective ascent and safety of the team over the potential dissent of a few. 

\subsection{Positionality}
Details of my entrée into the community are included in the Appendix. However, it is important to note that over the course of the ethnography, I became part of the moderation team—the struggles, successes, and failures shared in this paper are as much mine as theirs. While the majority of the moderation team are not American, the plurality Reddit users on Reddit are \cite{bianchi_2022}. Because content on Reddit is contributed and distributed by its userbase \cite{massanari2017gamergate} many of the events that impacted the moderation team in 2020 and early 2021 are US-centric. Readers should also be aware that my identity as a white non-American living and working in the United States undoubtedly influenced my experiences living through, moderating content related to, and recording and reflecting on each of the following controversies. 

Finally, it is important to note that I have written from a particular position of power: that of a community moderator. As Matias \cite{matias2019civic} notes, moderators ``serve three masters with whom they negotiate the idea of moderation: the platform, reddit participants, and other moderators.'' As this paper will discuss, moderators’ relationship to power is versatile and complex. Nonetheless, their particular occupation is the lens through which this paper is written and should be read as a complement to work that centers the perspectives of users who have been affected by moderation (e.g., \cite{jhaver2019did, jhaver2019does, gray2021we, marshall2021algorithmic, thach2022visible, haimson2021disproportionate, are2020instagram}).   

\section{Results}
This section contains three controversies, each describing how r/AskHistorians moderators navigated power by drawing on practices from alternative models. The first controversy describes a disagreement over a moderation decision when the content in question is borderline acceptable; the second describes two concurrent instances during which the moderation team collaborated to fight racism on Reddit; and the third describes interlocking controversies in which power, equity, visibility, and interpersonal conflict intersect.

\subsection{Disagreement: Tensions and expertise in decision making}

\begin{figure}
    \centering
    \includegraphics[width=1.0\textwidth]{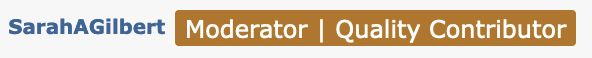}
    \caption{An example of flair on r/AskHistorians.}
    \label{fig:FlairExample_Unredacted}
    \Description{A screenshot of the author's flair. Next to her username, ``SarahAGilbert", is a brown box with white text that reads ``Moderator|Quality Contributor"}
\end{figure}

For the most part, decisions made regarding questions and answers shared on r/AskHistorians are made by individual moderators. They clearly violate the rules, often for being too short, a link to Wikipedia, or a snide comment, like ``Just google it''. However, borderline answers, such as those that demonstrate depth and at least minimal knowledge on the topic, are discussed by multiple members of the moderation community before a decision is made, especially when such an answer is written by one of the expert ``flairs'' (users who have demonstrated expertise in a given topic area and thus awarded a ``flair'' next to their username indicating their area of expertise. Fig. ~\ref{fig:FlairExample_Unredacted} shows a screenshot of my flair). Typically, these discussions are initiated by the moderator who spotted the problem. They, along with other available moderators, discuss whether or not the problem is severe enough to warrant removal, or if another form of intervention (such as a pointed follow-up question to elicit more depth) is more appropriate. This controversy shows conflicting roles of expertise in decision-making and why navigating these roles through labor of care is necessary for intersectional moderation. 

One of the more challenging discussions began when Becca brought an answer to be discussed in \#business-chat (bizchan). The question asked about university life during a time period that was part of her expertise. She stated that although it was answered by a flaired user, the answer was wrong because it was based on education in a later period. Thus, it misrepresented education at the point in time referenced in the question. She requested the answer be removed, and asked if we had a macro (a pre-written message available for all mods to use and in this case, used as a public form of moderation communication) that said, ``this answer is wrong.'' \footnote{Author: Reading the flair’s answer it was not clear to me how it was wrong. The answer acknowledged that little was known about education in the specific place and time and would instead discuss education at the same place but later time, an approach commonly used by moderators when the specifics of a question are impossible to answer due to gaps in the historical record. But I’m no expert so I felt like I should trust Becca. Yet, I was immediately uncomfortable. It felt as though there was no choice but to agree with her assessment and that bringing it to the group was just a formality.} Another moderator, Justin, responded to Becca with an alternative: rather than alert the flair through a public comment, could we not send the flair a private message? Becca declined Justin’s suggestion, explaining that she had a relationship with the flaired user and wanted to answer the question herself. Her response alluded to an internal policy in which moderators are not supposed to remove borderline answers to questions that they have answered or are intending to answer a question, as this creates conflict of interest in decision-making. Throughout the discussion, several moderators with expertise outside the subject area asked how egregious the error was and whether the answer could be salvaged with some feedback—another common approach taken by moderators when answers are on the cusp of acceptability. Becca explained that it couldn’t be improved because, in her opinion, there was too large a divide between education in the two time periods. As an expert, Becca was supported by other moderators in her request for a removal.\footnote{Aaron: While in this case, there is a chance that support for the removal was genuine, it's also realistic that support here came from a place of being too tired to argue. This was not the first situation like this. Disagreements about Becca's calls, especially pertaining to answers of flairs, had occurred several times before. While this one was particularly egregious, Becca had removed flair answers before without any sort of notification and taking arguments about it to a personal level afterwards. She was one of the longer standing members of the community who, also due to her PhD, writing and general high activity, carried a lot of weight within the community, which she would leverage in such cases, especially questioning why the rest of the team wouldn't trust her in her expertise. Not to overuse a Bourdieu-esque capital metaphor but her place in the community and group had led to the accumulation of a lot of ``capital'', which she converted to power in these cases.} While the decision to remove the answer was maintained, it was agreed to send a private message, written by Mollie. In the meantime, another flaired user answered the question. They also responded based on what was known about education in a later time period.

Several hours later, the mod team received a response from the flaired user whose answer was removed. In their message, they clarified that they meant to address the spirit of the question rather than the letter. Other moderators discussed the answer in light of this new information and decided that since the second answer was allowed to remain, this one should too. As the conversation continued, Becca rejoined the discussion to clarify her stance: while no information about education at that particular location exists, information about education in that era does. Therefore, the flair’s response was off topic. The decision was made to remove the other answer as well, and private messages were sent to the two flairs. \footnote{PH: Thinking back on the episode, one of the things that strikes me the most is that there was a strong pattern here, mostly (but I should stress not entirely exclusively) with Becca where it felt that we were expected to not only hold threads to a different standard when in their focus area, but even more so if they were thinking of answering the question, which went against the spirit of one of the most important guiding principles for us of ‘don’t mod where you post’, and thus was trying to be gotten around through pressure for us to mod it the way they wanted. This wasn’t the first time, nor the last, and this section in particular I think emphasizes how much deference we felt we had to give, as in a different topic, the kind of latitude we give for ``This isn’t the specifics of your question, but here is an answer on this very closely and obviously related aspect'' would have made this almost a non-issue to leave as is. It was essentially the conclusion reached eventually but only after hours of emotionally charged discussion and uncomfortable interactions with two flared users. Had it been a different moderator who raised this issue, I feel most of the team would have been much more comfortable in raising this point from the get go and nipping the entire thing in the bud before it even got off the ground, but the history of these discussions and knowing what would happen if you offered strong pushback to Becca meant no one wanted to invest in that kind of a response.}
 
After another few hours, another moderator, Keith, joined the conversation for the first time that day. He expressed that after reading the discussion he still didn’t understand the issue; if both flairs had interpreted the question the same way, shouldn’t that say something about how it could be answered? He and several moderators discussed the answer again and agreed that responses that address early education at that location were legitimate approaches to the question, even if the time period was wrong, and that both answers should therefore be allowed. Keith volunteered to draft two more messages to the flairs, letting them know that their answers had been reinstated. The second flaired user responded favorably, while the first didn’t respond.
 
When Becca returned to the discussion, she was not happy to find her earlier decision overruled; she felt that as the expert she should have been consulted. She followed up, providing more detail about the difference between the two answers, explaining that the second answer addressed the problem with the question (and was therefore acceptable) whereas first demonstrated a lack of knowledge and used inaccurate and older sources. Responses to Becca were mixed. Some were confused since she didn't note these issues before, while others supported her. Because Becca was the expert, their stance was that they didn’t need to understand because she was positioned to adjudicate with a nuanced perspective on the topic. However, the mods who didn’t feel like they needed to understand her reasoning also hadn't volunteered to send messages to the flairs.  

While the decision appeared to have been made, the next day the issue re-emerged when Becca brought it up again. She’d heard from another user that the flair had made a similarly egregious error on another answer (however, she didn’t share the answer). In response to her comment, Keith asked her what she thought we should do. Becca requested that because she had a relationship with the flair \footnote{Aaron: It cannot be overstated that this was part of a pattern. Whether she had a relationship with flairs whose answers she removed or not, she never wrote removal messages. Writing these messages is not a particularly easy or enjoyable task since it takes a certain kind of balance and tact so as to not discourage people from participating in the future. Writing these is not something that people, in my impression, particularly like and it is usually accepted form to write them yourself if you are an expert in said area. Also, it was my distinct impression that she would often remove stuff from flairs when she planned on answering a question.

PH: Aside from endorsing Aarons’s remarks here already, this was another aspect of the pattern remarked on before, as at least in several cases it was made clear that Becca did not want to be the one to write the message because of this personal relationship, either mentoring or apparently working on some sort of project. They were one of the most visible mods for flair outreach and trying to cultivate good relations with them, so it came off as specifically pawning off to others being the disciplinarian to maintain the outward appearance of ‘cool mod’. I would also say that it worked, as some of the flairs who had the best impression with Becca were ones she was incredibly harsh on at times privately, resulting in far worse impressions towards the mods who had outwardly communicated those issues} she would like someone else to write a private warning message. She also suggested that people start listening to experts, and mentioned that she was still upset about the prior day’s conversation. Keith asked her what she wanted the message to say, reminding her that he couldn’t write it on his own because he didn’t understand what the issue was. Erin volunteered to write the message and began to draft it. In doing so, the conversation shifted back to discerning the issue with the answer, which was still not clear to several of the moderators, in particular, why the second answer was fine, but the first was not. The discussion began to move quickly and questions, positions, and statements began to get lost in the fray. Becca reiterated her point, but it didn’t help because the team still did not understand what they needed to communicate. As the discussion progressed, Becca grew increasingly frustrated until she finally gave up, instructing the team to do what they wanted. In response, Mollie called out to her to wait, explaining that she was trying to listen but that she needed to understand the issue because she was often the one who responded to users:

\begin{displayquote}
\textit{
I want to be as supportive and helpful as possible when I tell them why their answer was removed - both for ethical reasons (I try to be good to people) and for logistical ones (flairs melting down in panelsub about how mean we are is a nightmare). Which does mean I tend to err on the side of saying their post is fixable, which I think tends to contribute to this repeated problem as it downplays the ``you don't know'' angle, but each time this happens I tend to be genuinely confused about whether you're saying it's fixable or not.}
\end{displayquote}

In response to Mollie, Becca reiterated that she didn’t think the answer was fixable. When another mod asked if the flair could adjust the comment, Becca provided a detailed breakdown of the issues. Mollie clarified that while she wasn't convinced by each of the arguments, that this was the most detail Becca had provided and that she had initially had different impression of the issue. Becca responded that she didn’t go into as much detail before because she hadn’t read the answer in its entirety; rather, she knew the post had to be removed right away. Based on Becca’s feedback, Erin drafted a message that was ultimately sent.\footnote{Erin: My incredibly gregarious grandfather was fond of telling all his grandchildren that if a person sees something that needs to be done and has the time and the ability to do that thing, and chooses to not do that thing, we were letting him down. That message, plus other socialization I likely got as a Gen Xer (I am, I believe, the oldest active mod), explains why I often volunteer(ed) to do things related to communication on the subreddit. In many cases, I had no real emotional investment before beginning the task. I just saw a thing that needed to be done, had the time and ability, and a fear of letting my grandfather down.}
 
The flair did not respond well to the message, saying that they were insulted. When another mod responded that no insult was intended, the flair followed up with more detail, writing they worked hard on the answer, it was positively received by those who had read it, and that its fate rested on the outcome of discussions to which they were not privy and could not contribute. While they initially enjoyed participating in the community, it now felt negative and anxiety-inducing.\footnote{Author: When I read the response my heart sank. Prior to this incident, most of the removals I had observed were obvious violations or borderline answers that had been successfully scaffolded. This was the first instance of an outright removal of an in-depth answer, and I hadn’t seen the time, care, and intellectual labor flairs expended and how that made them vulnerable to our power as moderators. I had felt confused and skeptical of decisions made before, but reading the response marked the first time I’d ever outright disagreed with a moderation decision made by the team.}

In the aftermath of the incident, a discussion was initiated on r/askhistoriansmods suggesting a policy change for managing flair removals: rather than removing a problematic answer outright, moderators could discuss it with the author. While it received support, the policy was never formally adopted. The initial argument would also resurface a week later and the tension would cause several moderators to leave bizchan to avoid the conflict. Two mods, Ethan and Aaron, engaged Becca in a lengthy discussion, during which she would explain that she was hurt because no one listened to her or valued her expertise. The others explained that the situation was stressful for the team because they didn’t understand the issue and therefore couldn't communicate why it was problematic to the author. By the end, very little was resolved. This argument, combined with others that were similar in nature, would continue and compound over the course of a year and a half.\footnote{Author: I would write in my notes how much it hurt to watch this discussion unfold. The fact that it resurfaced a week later felt destabilizing. After this incident, even when discussions appeared to be amicable, I couldn’t be sure that they were. I’ve never been one to hold grudges and I wasn’t sure how to operate in a space that seemed calm on the surface, but under which deep resentments would grow

PH: More a reflection on your reflection, but while I wouldn’t say discussions weren’t amicable, these incidents definitely built up, and of course impacted discussion and approaches even on issues raised with far less stakes. Especially with removal requests from them where it wasn’t immediately obvious what the issue was - beyond, as Aaron already raised, the suspicion that they wanted to answer the question - I know that I was not the only member of the team who essentially felt that before responding I needed to take a moment and consider what my emotional bandwidth was for the day, and I just wouldn’t speak up unless I thought I would be entirely in agreement on the issue, or really in the right headspace for a disagreement.}
 
In sum, this story highlights challenges assessing borderline content. Not only are decisions about what's okay and what's not difficult, assessing them requires expertise. The stakes are high in intersectional models as these decisions can decrease trust users have in the information they encounter. Finally, it shows how stressful decision-making can be. Moderators in these conflicts must choose between supporting their friends, who they respect, and relationships with regular users, who are in less powerful positions.  

\subsection{Collaboration: Banding together to fight racism on Reddit}

On May 25th, 2020 George Floyd was murdered by police. In response to yet another unprovoked murder of a Black person by authorities, Black Americans started a global movement to fight anti-Black racism. In response, Reddit-users began to ask questions related to current events. However, rather than ask questions related to the history of racism in the United States, they asked questions in ways that centered white histories or othered Black protesters. For example, they asked questions about responses to looters during the LA riots, the effectiveness of revolutions and resistance movements in largely white Western countries, and terrorism. The team also observed a rise of users asking questions with the intent to promote racism while maintaining plausible deniability, a tactic known as ``Just Asking Questions'' \cite{breit2018how}.

Within a week of Floyd’s murder, large corporations began making statements professing their commitment to anti-racism, including Reddit. On its blog, Reddit CEO Steve Huffman wrote, ``As Snoos, we do not tolerate hate, racism, and violence, and while we have work to do to fight these on our platform, our values are clear'' \cite{huffman2020as}. His words contrasted with earlier statements by Huffman \cite{huffman2018} and Reddit’s longstanding resistance to banning racism on the platform \cite{chen2012reddit}. Reddit users’ information-seeking patterns and Reddit’s stance on racism drove the r/AskHistorians moderation team to take action through two overlapping movements. This is a story of how creative and collaborative moderation labor can drive social and platform justice.
 
On June 2, 2020 Allison brought an idea to \#business-channel: a former member of the modteam had suggested putting together a formal statement on police violence. As of yet, the topic hadn’t organically arisen on the subreddit through questions, and so it immediately gained support. The team excitedly began to brainstorm what the post should include, with several moderators offering to help write and edit it. However, few of the current members of the moderation team had specific expertise in anti-Black racism and none were Black. While there was some discussion of reaching out to members of the flair team who may have such expertise, this was ultimately dismissed, partly due to a sense of urgency. The initial brainstorming for the post was collaborative. Allison took on a leadership role and together, the team decided to write an historical overview of police violence in America.  
 
As brainstorming continued, news of Reddit’s statement reached the team and everyone was in agreement: the statement rang hollow, as the platform continued to host racist content and communities. Several hours later, shortly after 5:15pm, Edward entered \#business-channel to share some news. He had learned that several subreddits, most notably r/nfl, were planning to ``go dark'' (i.e., restrict participation in their communities by setting them to ``private'') to protest Reddit’s ineffective response to racism on the platform. He asked, ``is this something we want to follow suit on?'' As with the post on the history of police violence, there was immediate support for participating in the protest. However, rather than go dark, the team decided to restrict participation in order to more effectively convey a message since going dark would limit the message to a short blurb on the splash page and thus would only be visible to people directly navigating to the subreddit’s homepage. On the other hand, as PH noted, restricting participation meant that content on the subreddit could still be read and voted on, including a longer post explaining the rationale for the protest. Theoretically, he noted, the message could hit r/all, (the ``front page of Reddit'' consisting of the most highly upvoted posts from across the site and visible to all Reddit users, including those without accounts), reach a broader audience, and have greater impact. Further, there was some concern that going private could alienate users. Ashley and Allison, with input from others, drafted a message to communicate the protest. At 8:30pm, the protest message was posted.
 
Almost immediately, it was clear the protest post would be popular. Within 3 minutes of posting, it was the 4th ranked post on the subreddit’s homepage; in the first hour it would have over 1000 upvotes; by 11:30pm it reached the front page of Reddit, peaking at \#4 on r/all. In total, the post would receive 33,933 upvotes. However, it also drew ire. Within 12 minutes of posting it would receive the first report (for soapboxing). By the time the post would run its course, it would be reported approximately 112 times (some would be removed by Reddit’s administrators’ and so the exact number of reports is unknown). Of the remaining reports, the majority reported it violated the subreddit’s rule for engaging in soapboxing, current politics, or recent events (31), followed by misinformation (27) and spam (19). Twelve reports were custom written by users. Since reports on Reddit are anonymous, the reporting system is often used to send abusive messages to moderators without risk of sanction. Among the open text reports were messages containing racism, antisemitism, gendered slurs, and ableism. By the following day, 98 subreddits would message r/AskHistorians expressing solidarity, 90 of which also shut down in protest while 8 communities expressed solidarity but could not shut down as they provided important sites of support or organization (e.g., r/washingtondc and r/BlackLivesMatter). The movement was picked up by the popular press as well. Newsweek featured r/AskHistorians’ position, which was collaboratively written by the moderation team \cite{Asarch2020reddit}.
 
Because they could not comment in the thread or elsewhere on the subreddit, people began communicating with the moderation team through modmail. The messages moderators received ranged from supportive to abusive and bigoted. They also began messaging individual moderators, whose moderation status would not be visible had the subreddit been set to private. The moderator who posted later noted that as of the following morning, they had 62 inbox messages and 11 chat request notifications. One of the moderators at the top of the list of moderators remarked how many racist messages they had been getting.
 
Meanwhile, the team discussed the reports as the modmails flooded in and continued to collaborate on the post about the history of police violence. Allison, who was drafting the post, did not sleep that night. When she mentioned that it was making her anxious, Liam offered to take a look. Allison explained that she was having trouble moving beyond providing a list of examples to the central thesis. She wanted to ensure it was clear that police brutality against Black Americans was central to policing, resulting in a system of white supremacy that benefited white Americans. Allison wasn’t the only team member feeling anxious. The closed subreddit meant that there was little to do but wait until it was opened back up again. Liam, in his efforts, struggled to keep his addition to the post aligned with community norms \footnote{Liam: It was particularly difficult to find the right tone, words, and content because I am not an American and while as a European, there is the inevitable cultural reception of US-American issues with racism, police violence and white supremacy. Even more than the white Americans on the team, it is a topic that I can speak to only with a rather massive distance and so I felt an insurmountable pressure to find things to say that went beyond a general indictment of what I understand as white surpemacists culture in the US.}. In response, he and Brandon engaged in literal gallows humour, joking about guillotines. Matt and Michael joked about their anxiety—Matt joking about his failure to drink coffee without spilling it while Michael made puns. Allison, still struggling with the post, also joined in the joking, as did Jason. When she asked for help finalizing the post, Michael provided suggestions for framing the introduction and conclusion, PH shared a list of relevant sources, and Jane helped copyedit. At just after 9am on June 3, the team decided the post was ready to share.
 
When the post was shared the team decided to allow comments, although the community would still be closed for submissions for several hours. While rules for commenting on the post were somewhat relaxed from the normally very strict rules, the team monitored it closely. In less than 20 minutes the post attracted its first racist comment, which was immediately removed. As more comments came in, mods began sharing examples from the thread to discuss and (in some cases) joke about them. For example, Matt shared this removed comment saying, ``It’s white Fragility: The Post'':

\begin{displayquote}
\textit{I used to enjoy this sub. How dare you place the blame on me, a 33 year old from 2000 miles away from where this happened? Seriously, go fuck yourself for that, you asshole. You honestly think you’re helping anything by doing that? You’re not, genius. I wasn’t even alive for 90\% of what you describe. I refuse to let you make me feel guilty for things I’ve never done. I honestly thought the mods of this sub were better than posting hyperbolic nonsense titles. I’m done...}
\end{displayquote}

The post also attracted complaints from users for being too political and violating one of the subreddit’s stated rules, ``Nothing Less Than 20 Years Old, and Don't Soapbox.'' These comments were addressed several times in the thread by moderators who noted that the mission statement of the community includes advocacy. In private discussions, they would note that not making a statement is also political, yet commenters did not question that.  
 
After being up for about four and a half hours, the post reached the 11th spot on r/all. The high spot on Reddit’s ``front page'' increased the traffic coming to the post. In addition to comments in which users took personal offense, the thread also began to see an influx of hate-based disinformation, primarily in the form of false, misleading, or incomplete statistics; links to racist blogs and misinformation websites; and complaints that the post centered race over class. While comments containing overt racism were immediately removed by moderators, moderators were challenged by when to use bans. Aside from the most egregious cases of racism, in which banning was an obvious and immediate response, which users should be banned, warned though a temporary ban, or simply ignored? Matt, who had been actively moderating the thread for hours, noted that some of the commenters, particularly the ``facts-aren’t racist'' types, were not necessarily white nationalists but rather people who don’t think critically. He asked if banning was too harsh a punishment for them. Aaron noted that if they were coming to the thread to complain, they likely didn’t have good intentions. At that point, it was agreed that comments such as those would result in a ban. Comments that centered class were removed without banning because it was determined they derailed the conversation, weren't made in good faith, and thus exhibited racism. These comments were a source of immense frustration for members of the moderator team, in part due to their plausible deniability; several such comments had even accrued upvotes and awards before they were removed.
 
When users are banned on Reddit, they can reply to the ban via modmail. Ban messages were often responded to with even more egregious racism and aggressive comments instructing the mod team to ``kill yourselves.'' They also referred to the team as ``janitors'' or ``jannies,'' terms commonly used for moderators on free speech and right-wing platforms and intended as an insult. For example, one user sent three rapid fire modmail messages, writing:

\begin{displayquote}
\textit{Slandering cops and forks [sic] talking about civility? 
Stfu cuck mods}  

\textit{You’re an online janitor working for free bitch}

\textit{Stop slandering police you worthless degenerates. [Redacted statement containing homophobia and anti-Black racism].}
\end{displayquote}

A report would read: \textit{``clean it up janny, clean up my shit FOR FREE you pathetic janny.''}

In response to messages such as these, the modteam began joking. For example, PH commented that ``We do it for the warm fuzzy feeling we get we we [sic] banned you,'' and Oliver described it as, ``I look at it like barter or gardening. I get way more value out of the sub, then effort paid putting into it.''
 
\begin{figure}
    \centering
    \includegraphics[width=1.0\textwidth]{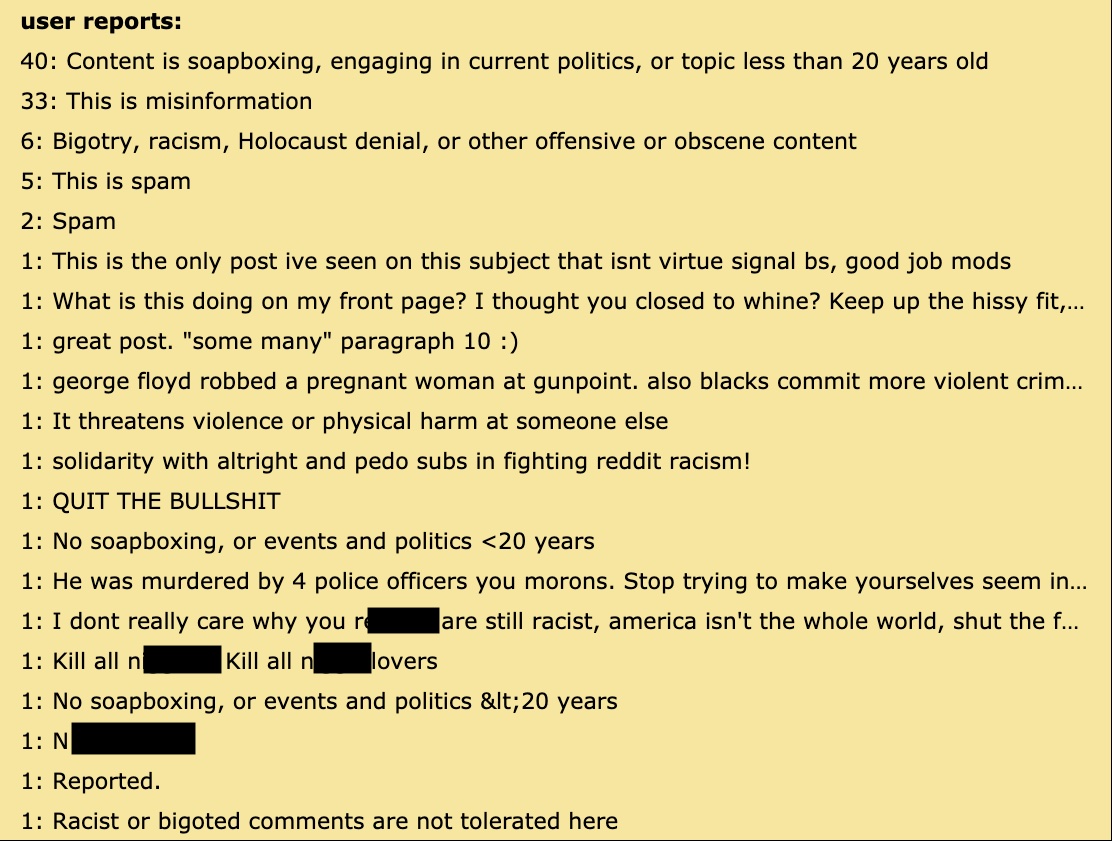}
    \caption{Reports on the History of Police Violence post. Several of the reports visible in the figure would later be removed by Reddit's administrators.}
    \label{fig:RacistReportsRedacted_copy}
    \Description{A screenshot of reports made by users on the History of Policing thread. The number indicates how many times each report was made. Spelling and grammar are retrained from original but slurs are redacted. Some report reasons are cut off. The text reads: 40: Content is soapboxing, engaging in current politics, or topic less than 20 years old; 30: This is misinformation; 6: Bigotry, racism, Holocaust denial, or other offensive or obscene content; 5: This is spam; 2: Spam; 1: This is the only post ive seen on this subject that isnt virtue signal bs, good job mods; 1: What is this doing on my front page? I thought you closed to whine? Keep up the hissy fit, . . . 1: great post. ``some many'' paragraph 10 :); 1: gerorge floyd robbed a pregnant woman at gunpoint. also blacks commit more violent crim. . . 1: It threatens violence or physical harm at someone else; 1: solidarity with alright and pedo subs in fighting reddit racism!; 1: QUIT THE BULLSHIT; 1: No soapboxing, or events and politics ,20 years; 1: He was murdered by4 police officers you morons. Stop trying to make yourselves seem in. . .; 1: I don't really care why you r-slurs are still racist, america isn't the whole world, shut the f. . .; 1: Kill all n-slur. Kill all n-slurlovers; 1: No soapboxing, or events and politics \&lt;20 years; 1: n-slur; 1: Reported; 1: Racist or bigoted comments are not tolerated here}
\end{figure}

As with the protest post, this one also accrued numerous reports. While the most egregious reports were removed by Reddit’s administrators once they were reported, 136 reports remained: 52 reports for misinformation, 51 for soapboxing, 11 for spam, and 7 for racism. The remaining were open-ended responses, most containing some form of racist, ableist, and homophobic speech. Figure ~\ref{fig:RacistReportsRedacted_copy} shows a screenshot of the reports taken June 3, 2020 at approximately 6:30pm. As can be seen from the numbers, the post would accrue more reports after the screenshot was taken.  

While the post attracted considerable pushback, it received far more support. All told, it would amass 38,234 upvotes, which made it the most upvoted post on the subreddit of all time. There were more positive comments than negative, and the positive comments were frequently upvoted. People came to the thread to share additional information about police violence and its causes (such as slavery, redlining, and racist educational practices); share Black histories that they had recently learned; recommend books and articles; ask follow up questions; and thank the moderators for writing the post. 

Shortly before 8pm on June 3rd, Reddit posted a response to the wave of subreddit protests. It stated that its administrators were planning to reach out to moderators impacted by recent events and create a new Community Council. They mentioned that they would update mods regularly about developing solutions, and would continue to expand the Community Council Program. The r/AskHistorians mod team was skeptical that any progress would be made, particularly as the actions set forth in the post were vague.\footnote{Author: I was incredibly frustrated by what I viewed as a milquetoast response. So much so that I put my energy into responding to the post in a (slightly) snarky comment. Initially, responses to the comment were positive. However, after it was linked to two right-leaning meta subs (subreddits for discussing reddit) I began to receive pushback in the post and several abusive private messages. That night I double checked that my personal social media accounts were set to private and kept an eye on my Twitter account; fortunately, the abuse was limited to Reddit.} On June 4th, a more concrete action was taken: a ``yikes'' award used to abuse users was removed by the admins. However, they did not remove a different award that was shaped like a monkey, despite moderators of r/BlackPeopleTwitter explaining that it was used on posts in their community for racist purposes. On June 5th, Steve Huffman posted on r/announcements, listing concrete changes Reddit planned to make, including updating their content policy to explicitly address hate.\footnote{Author: While I had not done much to help write the posts, I felt very proud to be part of the r/AskHistorians moderation team in that moment. I had been feeling incredibly helpless as I’d wanted to join the protests in my city but was anxious about arrests and the impact that could have on my ability to stay and work in the country as a non-resident alien. Removing racist comments, banning racists, collaborating on responses to the press, and writing snarky comments all felt like I was finally able to do something, even if it was small.}
 
The moderators' posts were a success—a product of the team collaboratively wielding proactive moderation, and using the power of the community and its influential position on Reddit to work with other communities to pressure the site into a response. They provided historical context for current events to Reddit users, and prevented the promotion of hate by preemptively removing racism and dis/misinformation. Ultimately though, moments of coming together such as this and others would provide temporary reprieves.\footnote{PH: I would go further even and say it wasn’t merely a reprieve at times, but sometimes quite specifically an excuse for us to shoddily patch things over. I can think of at least one case where there had been some very emotional and divisive discussion of an issue similar to the one raised earlier, only to be interrupted by Jan. 6th, which resulted in a very high profile and active Feature thread. It had that ‘bringing together’ effect, and pretty specifically stopped that discussion from continuing. The few days ‘pause’ that happened killed the momentum, and probably for the worse as the issue came back even stronger some months later.} While shared values would bring the team together towards common goals momentarily, they would not address the team’s underlying tensions.

\subsection{Conflict: Anyone who loves communities or sausages should never watch either being made}

Conflicts on the moderation team could fester beneath the surface for weeks, months, or even years before erupting again. These eruptions felt difficult to predict, as what seemed mundane to some could be sacred to others. However, they often coincided with particularly challenging moderation decisions. This is a story of emergent challenges from missing context, and attempting fairness and transparency in decision-making. These challenges were exacerbated by longstanding and often invisible interpersonal tensions.

In one sense, this story begins on Feb 14, 2021 when Andy, a former moderator, came to bizchan to make a request of the mod team. In another, it begins years before that, when Andy had been hurt by the team. Since he seemed to have left the team in good standing and was still active on Slack, the newer members of the team were unaware that he was hurt and many of the older moderators believed he had healed. While Andy’s hurt was hidden, other tensions on the team were clear. Factions had begun to form around two moderators, Leah and Claire. Leah tended to take a strict approach to moderation, prompting concerns among some on the team about ``standards creep'' creating high barriers for participation. Claire, once a close friend of Leah (and one of the few women on the team), was often the one who would confront Leah during disagreements, prompting some on the team to think of her as a bully. These longstanding conflicts (along with elements of Slack’s design) would influence how the following events would play out.
 
It was late in the evening on Feb 14 when Andy joined bizchan. A discussion with a user on the subreddit had gotten heated. The user accused a leading (African Caribbean) historian of being a white supremacist and Andy of the same, and so he asked if the comment could be removed as a civility violation. After making the request he left the channel. Andy’s exit and entry in Slack was made visible by the platform, which announced when he joined and again when he left. Similar quick entrances and exits, while not a common occurrence, were sometimes done by former moderators who were still active participants in Slack’s social spaces when they wanted to highlight a moderation issue. The comment Andy highlighted as a civility violation was removed immediately and the discussion shifted to further sanctions: a verbal warning, a temporary ban, or a permanent ban. The user had posted a few high-quality answers in the past, but a deeper reading of the comment suggested that some type of ban would be appropriate. Most involved in the discussion supported a temporary ban, although a minority argued for a permanent ban, as that level of hostility directed towards a regular user would not be tolerated. Claire, one of the moderators supporting a temporary ban, outlined her rationale. While she agreed they hadn’t responded with civility, she was hesitant to levy a permanent ban based on the user’s post history—they appeared to be a person of colour and she did not want to ban someone for expressing anger at what they perceived to be racism. PH issued a temporary ban, and within minutes, commuted it to a permanent ban due to the user’s hostile response in modmail.\footnote{PH: One thing that always struck me about this, and how it relates to the larger issues of gender and institutional issues there, is that despite being the one who actually wrote the message, I don’t think anyone ever said anything directly at me. Claire took all the brunt of it. Maybe it was simply because I had some real life stuff going on so wasn’t online for the next few hours, which meant Claire was the main one who articulated the position, and other factors can also be explored, but certainly even after I returned to the conversation later in support, the only mod who took blowback seemed to be the woman, despite several men holding the same position.}

Although the user had been sanctioned as a result of the interaction, the discussion continued when Leah noted that she was disappointed in the decision that was made. She explained that the personal attack was in public and could be seen even after it had been removed through websites that display moderator-removed comments. Claire explained her original argument, saying that while she felt bad for Andy, it felt wrong to ban someone because a white person felt hurt when being called a racist. She described this kind of ``civility'' breach as fundamentally different than the other kinds that the team routinely deals with, stating: 

\begin{displayquote}
\textit{if we are actually to be inclusive and a safer space, we absolutely need to not treat accusations of racism, homophobia, misogyny, etc as insults and not heavily weight how much said accusations hurt the privileged people accused as part of our decision of how to handle the situation.} 
\end{displayquote}

Viktoria chimed in to disagree with Claire. Because of Reddit’s primarily text-based interactions and pseudonymous participation, she claimed that all we could rely on were self-descriptors, which could not be trusted.\footnote{Viktoria: Anonymity on reddit often means anyone can claim to be anyone. However, the default tends to be cis, straight, white male until proven otherwise. Being that interactions are often singular points of contact means that a dishonest member can claim to be a marginalized person, be an expert in a field, in an effort to shut down discussion or ``win'' an argument and give credence to their claims and then choose another identity when the opportunity arises. This anonymity can however be used by marginalized people to shield themselves from harassment due to the assumption of the generic identity by other members. When discussing the idea of using profile histories as some indicator, it raised a red flag to me, as a trans woman, because if you used my profile history there would be no indication of my status. This has been a conscious effort by me to hide my identity and blend with the assumption by most reddit users. If I was more forward, it could lead to harassment, doxxing of my private life, and hatred directed at me and the subreddit as a whole. This hidden minority, could stretch farther and wider than generally understood and would lead to a lot of mistakes being made. It could also lead to the ``hidden'' minority to feel compelled to out themselves to our team, which does not feel like a policy we should enact.} Further, she argued that moderating this way could be viewed as patronizing. As it was late at night, Claire left the conversation, stating that she disagreed but was going to bed. Leah and Viktoria continued to discuss the issue, noting then that the user had a history of posting in a number of discriminatory communities and making contradictory claims, including about their own race.
 
Later that night Andy sent a modmail. He had seen the conversation, having been alerted by username pings. Having viewed it, he was hurt by the discussion he witnessed. Reacting to Andy’s modmail in Slack, one mod noted that he was disappointed. While he didn’t know Andy well, he seemed nice and was upset that the team had hurt him by opting to take a lighter moderation approach ``because of some `diversity' intiative [sic] based on a five second read of a profile'' (he would later apologize after several moderators of colour explained that they had been hurt by the statement).

The next morning the team discussed how they could best address Andy. In the middle of the night, Claire had woken up and written a long comment about how hurt she was by the discussion that had continued without her. The incident, and others, had made her consider leaving the mod team. However, only a few moderators had acknowledged it at the time and so later she re-upped her comment from the night before. She stated that while she understood why it hadn't attracted people’s attention, she needed the issues she noted addressed to stay on the team.\footnote{Author: Waking up to the discussion I was deeply frustrated and confused by the tone of the conversation and the focus on Andy. Even after reading his modmail I felt as though his response to the discussion was unfair and lacking in appropriate context. I was upset that he had violated a space the team considered private, and then lied to us about it (I was certain, and it would later be confirmed, that naming someone without @-ing them in a channel they were not part of would not cause a ping). Further, I didn’t understand why the focus was on Andy, a former moderator, and not on Claire, a current, and also very hurt, mod.} Re-upping it drew support from several moderators, who agreed to post on modsub to share their understanding of what happened the night before. Lily volunteered to write the post. The day before, prior to the conflict stemming from Andy’s complaint, I had brought up racism in algorithmic moderation. Since Claire’s comments had been met with some pushback, I believed this would be a good time to bring up how moderation can perpetuate racism. When the discussion continued to focus on how the argument played out with Andy, I left the channel.\footnote{Author: Truth be told, I was hopping mad and I wanted everyone to know. While Claire may have missed contextual clues about that particular user, I felt strongly about the issue. I was frustrated that Andy was centering himself instead of recognizing that the discussion that had upset him was an attempt to create a fairer space, and that his narrative was the one framing the discussion. So when I attempted to bring the discussion back to what I viewed as the larger issue and the conversation kept getting redirected back to Andy, I saw red and figuratively stormed out of the room.} In response, the mods called for a pause on further discussion until the team heard back from Andy.

Andy sent a reply several hours later about why he was hurt, and that our initial response exacerbated it. In short, the initial response hadn’t engaged with his concerns and had come across as face-saving rather than a genuine apology. Based on his experiences with the team, it was a response he had unfortunately come to expect.\footnote{Kevin: This might be something that Andy would have said was exactly the problem but to this day, I am unclear what sort of response he would have liked or what we could have said to alleviate his hurt in this case. By now I think there is indeed little we could have actually said that would have caused Andy to walk away from this without anger. In my impression this was the result of a long festering anger stemming from past conflicts that were never really brought up again or resolved that erupted here.} The moderator who had written the response, Lily, did not know Andy well. Those who did later noted that, given the context of issues Andy had with the team, they were not surprised he didn’t respond well to the message. The next several hours of discussion would focus on how to address his most recent message and a different moderator sent another message. Later that day, Andy would respond again, this time with a link to a 45-minute-long audio recording. When one moderator asked for a summary of the recording since he couldn’t listen to it. Another, Lily, who was currently listening to it, reported back: he accepts our apologies, he understands that he saw something out of context, it was a bad series of coincidences. He’s doing well and is in a good place, but he’s going to take a break from the community and Slack. The team was relieved by the news that common ground was found and they were happy to hear he was doing well. There was more, but since Lily didn’t understand the context or its significance, she didn’t share it in her summary.\footnote{Lily: It was obvious to me as a listener that Andy was working through several deeply personal matters as he talked. I knew that a number of mods were feeling anxious about what was happening and my goal when summarizing was to alleviate some of those concerns if his message allowed for that. I remember feeling relieved when I found things I felt comfortable sharing.}

In the meantime, the thread posted on modsub was not going well. Several moderators, including me, noted that they had witnessed and experienced identity-based inequity when interacting with other team members. While some of these contributions were met with support (such as mine), one thread, written by Beth (the only woman of colour on the team), was derailed when one mod suggested that moderator seniority be accounted for when considering intersectionality within the mod team. Several moderators pushed back on that statement, noting that senority is not related to identity and therefore irrelevant to the discussion. In another thread, Jim focused on the overall toxicity of the team, caused in part by its unwillingness to trust the expertise of other members on the team, indirectly referencing recent conflicts between Claire and Leah.\footnote{PH: I think it is also worth noting here that Jim had posted a thread a little over a month previously on essentially the same topic, and largely done in what he saw as a defense of Leah. A number of moderators responded however and none of them agreed with how Jim had characterized the issue. The thread did not go at all as he had expected, and he replied to almost no one, basically checking out of the discussion which died and led nowhere. So while long simmering, this issue was also very fresh in that regards, and I would also note it was one that Jim essentially knew how everyone would react to, but wasn’t actually willing to discuss.} When Lily explained why his desire to comment on moderation disagreements between two women (and often regarding women’s history) was problematic, Jim responded that the team was split into two ``Mean Girls'' cliques, and that Lily was not above the fray. When she asked what she could do differently in conflicts, Jim told her to ``turn off the schoolteacher-trying-to-draw-out-the-correct-answer-from-the-recalcitrant-student tone.'' An expert in the history of education, his response hit Lily hard: ``I'm not sure what you were hoping to accomplish by sending that to me but it's a fairly effective one-two punch.''\footnote{Lily: Figuring out how to be a good human on the internet, much less Reddit, is no easy feat. Trying to do that as a white woman of a certain age – knowingly interacting with younger people, mostly cis men – can create some fairly remarkable cognitive dissonance around balancing expertise with perceived bitchiness. What made Jim’s remark sting so much was not only was it unnecessarily mean, it hit at one of my soft spots. For good or bad, for right or wrong, I do try to approach online interactions as conversations with the goal of understanding where the other person is coming from. To be sure, I wrote and deleted a number of equally mean profanity-laden comments but I hoped articulating how his comment felt to me would have a greater impact.} As Lily disengaged, PH responded to Jim saying that in a thread discussing unhealthy gender dynamics, an insult that has gendered implications was highly inappropriate.

After this exchange, Jim quietly removed himself as a moderator. Whether or not anyone noticed, his departure would not be discussed by the team for another few days. He was the first of three moderators r/AskHistorians would lose over the course of the week. The second would be Sean, who had been quiet throughout this argument. Yet, as he would state in his resignation post, he would find himself glued to Slack during arguments. It was too difficult for him and he had to leave. Shortly after the conflict between Lily and Jim, Leah started a new thread asking for an indefinite pause on the discussion.\footnote{Kevin: This was a highly unusual request. Discussions, especially difficult ones, tend to peter out due to the nature of our mediums of communication. To put a moratorium on a topic is usually not done and to me seemed like an attempt to control the conversation rather than calm or resolve it.

Lily: The only time I seriously considered walking away as a moderator of AH was immediately following the posting of her request. I was livid. Partially because, as Kevin describes, it was highly unusual. But also, I had invested what felt like a huge amount of time in PMs with Leah, talking through various issues, working through tensions, and advocating for resolution. Her post made it feel like I had wasted my time.} As with the other post, this one would also not go well when moderators reacted angrily to her request to pause what they viewed as an important discussion. Discussions on this thread would span several days and prompted the departure of the third moderator, Dave. On February 19, Dave used that thread to attack Claire because he saw her as the cause of the team’s conflicts. To ensure his comment was seen, he copied the link into Slack and used the @channel command to alert everyone in bizchan. He then continued to attack her in Slack. As the conflict heightened, Kevin—who had been selected as ``Snoopy'' (a rotating position designed to move projects forward, lead new initiatives, and manage conflict)—removed Dave from Slack and as a moderator in an unprecedented intervention that the team would later (nearly unanimously) vote to uphold.\footnote{Author: I’ve omitted the details that prompted Dave’s demodding because they’re too sensitive. But the incident, only 45 minutes long, was easily the worst thing I experienced as a moderator. In my notes, I laid out a plan to leave the team if Dave returned. To put it lightly, I was relieved when his demodding was upheld. 

Kevin: This was easily the most stressful thing I ever did in connection to AskHistorians. In the past there were users who messaged me about dismembering me and due to my past posts there is always the danger of Nazis trying to dox me but that has never induced a level of stress comparable to this. I initiated what was a massive breach of the underlying trust of the community and that violated an implicit social contract among the mod team (due to reddit's architecture more senior mods can remove less senior mods–demodding less senior colleagues is a potential Damocles sword hanging over a community). And I didn't do this lightly. I genuinely felt that especially in light of mounting tensions of previous months and a general decline of good will between us, this had the potential to completely blow up the entire team. Before my inner eye I saw a mass exodus leaving the sub virtually unmoderated and so I did what I did and removed Dave.} 

The morning after the vote ratifying Dave’s demodding, discussions on Slack had almost returned to business as usual. Avery wrote a proposal to form a Diversity and Inclusion Oversight (DIO) Committee and the vote quickly passed. A question about communism prompted a slew of Marx puns.\footnote{Nick: A dynamic highlighted here and elsewhere is the use of humour - often puns - to break tension and move past particular incidents or deal with stressful user interactions. This is hardly unique to the modteam, but reads interestingly in these cases (as the choice of when to leap on an opportunity for humour can be taken as an effort to draw a line under an argument or incident, perhaps prematurely). I also think the tendency to make fun of users as a coping mechanism did feed into the more legitimate end of Andy’s complaints about modteam dynamics, with jokes at others’ expense building and reinforcing in/out dynamics that feed into other hierarchies (we, the mods, laugh at you, the user who asked yet another ****ing question about Hitler). Where to draw the line between acceptable/unacceptable targets and timing is hard.} After demodding Dave, Kevin stepped down in his role as Snoopy so PH, as the new Snoopy, drafted a message to Dave to let him know that the team had voted to uphold his demodding. Then, mid-morning,\footnote{Eric: It’s worth mentioning that mid-morning in the United States is during the night for the Pacific Rim mods, which often meant waking up to 300-600 messages during these events, some of them quite long, that I needed to try and get through in the hour between waking up and going to work. Another 300-600 messages could be posted during my work hours, which I often couldn’t read due to my job. It made keeping up with events as they unfolded very difficult and stressful, and I know that I frequently missed important context for understanding what was going on or had points I wanted to make but no longer could because the conversation had moved on two or three hours before. It sometimes felt very alienating to be mostly relegated to the sidelines in all these events because my timezone and work schedule largely prevented my participation, while at the same time I felt I was letting the team down by not being available to, for instance, help moderate the protest post during the peak periods.} Ben entered bizchan to ask if anyone else had gotten added to a private subreddit called r/askhistoriansmeta. No one had, and Ben clarified why he asked: Andy had used a subreddit that he'd created years ago and made private to publish an open letter to flaired users, who he had been manually adding to the still-private subreddit. The team discussed whether or not it would be ethical to share the letter with others on the team, ultimately deciding that it would be when another member of the moderation team was added.
 
The letter was long. It contained Andy’s concerns about the mod team, which he characterized as classist, homophobic, and ableist. While the team had experienced conflict and identity-based inequity, most stated that they did not recognize the team Andy described in the letter. Indeed, Andy had left the moderation team about a year before writing the letter, after having been mostly inactive as a moderator prior to that. Most of the moderators on the team had never worked with Andy and several had never interacted with him at all. Ian, one of Andy’s closest friends on the team, knew that Andy had concerns about homophobia. He had been portrayed as emotional and sensitive in the recent discussions about him, and Leah, who noted that she had also discussed this with Andy, agreed it was an issue.\footnote{Author: I read, and re-read the Slack transcript. I hadn’t noticed homophobia at the time and I wanted to see what had gone wrong so I could learn and do better. But I couldn’t find anyone describing Andy at all, let alone describing him as emotional or sensitive. I have no doubt that Andy experienced this type of homophobia while a member of the moderation team; I just don’t know how (or if) it manifested in discussing his actions that week which is why it's not noted in the story.}
 
It was also clear that someone on the team had been sharing the team’s discussions with him and that there had been some miscommunication; for example, the letter indicated that someone had been assigned to listen and report back on the audio recording, although Lily had volunteered. Then Ian revealed that he knew that Andy had been talking to someone, and that someone had been sharing details about the conversations, but asked that the someone reveal themselves. They never did. Later, in response to a user asking what proof he had that these discussions had occurred, Andy provided a redacted screenshot of part of the summary of the audio-recording. The breach in privacy left the team feeling vulnerable, especially the two moderators whose conversation was included in the screenshot. Technical fixes, such as locking the channel, no longer seemed sufficient if someone was reporting back to Andy and even sending him screenshots. In response, some team members began addressing Andy and The Leaker directly. Others noted that the sanctity of the space had been eroded and that they didn’t feel safe as long as The Leaker was at large.\footnote{Lily: Looking back on these interactions, it becomes clear that consciously or not, through formal and infomral means, team communication has shifted. We’ve collectively gotten much better at threading on Slack, we’ve created new channels for especially charged topics - the dedicated \#COVID19 channel has been a balm during the pandemic, and we’ve added details about our openness to PMs to our bios.}
 
The newer mods on the team were affected by Andy’s open letter differently than the mods who knew him. While they wanted to ensure they weren’t perpetuating identity-based discrimination, they also felt surprised by what had happened and that they were being held accountable for conflicts they hadn't been involved in. When Julie wrote a message expressing her frustration and dismay that the community she knew and loved and felt safe in suddenly seemed like it might be very different, a series of new moderators chimed in, saying they felt similarly. In response, Kevin—in collaboration with some of the older moderators—offered to write summaries of major conflicts, so that newer moderators would not be left in the dark. The older mods, several of whom considered Andy a friend, felt betrayed.\footnote{PH: I know that I was one of several who felt that they had been close with Andy to one degree or other, and who had reached out privately during all of this. I didn’t feel we as a team had necessarily handled it wrong, but certainly it pained me he felt hurt and I wanted to do all I could to help and mend bridges. It had resulted in an emotional (in the good way!) conversation on both sides, and had meant a lot to me. But that morning, while spending an hour or so on a reply to his (very kind) last message that I’d woken up to, \textit{that} dropped, which entirely changed how I viewed the conversation we’d been having, as clearly he’d been writing and planning it while talking with me. And while perhaps it meant I wasn’t specifically intended as one of the ‘problems’ called out in it, it nevertheless felt so absolutely devastating and was hard not to read the entire interaction as duplicitous. It also made it hard to presume that this wasn’t the inevitable endpoint from very early on in the episode no matter what we had done as a group.} What they’d thought was friendship was hard to reconcile with what felt like a malicious attempt to destroy not just the mod team, but r/AskHistorians itself.

Eventually the conversation in the private subreddit slowed and Andy stopped adding flairs before he made it through the list. The mod team made a post on the official private subreddit for flairs to be as transparent as possible. For example, members shared an unedited log of Slack discussions Andy had referenced in his letter. The next day, a bot was set up on Slack so that bizchan could be locked without losing access to discussion logs. Charlie, a long-time and highly trusted moderator, came off hiatus to investigate the breakdown in security. After several weeks of investigation, Charlie learned that while Andy had been talking with at least 10 moderators following the February 14 incident, no one had leaked screenshots from Slack. Because Andy had not been removed from the Slack space, as the team had still hoped to rebuild a relationship with him, he was able to preview discussions on bizchan without actually entering the channel, up until he had left the Slack space shortly before publishing the open letter. He had taken the screenshots himself. However, an open question remained: someone had shared the initial recap discussion on modsub—a space Andy did not have access to. While individual moderators developed theories about the likely culprit, the team, led by Charlie, collectively agreed that continuing the investigation to find a scapegoat would do more harm than good.

This controversy highlights challenges associated with enacting intersectional moderation. For example, to infuse moderation with fairness and justice, moderators need information that is not always available or reliable, and engage in deliberative processes that can be stressful to engage in and to witness. These challenges can fuel interpersonal tensions and exacerbate stress. While over time interpersonal tensions would ease for r/AskHistorians moderators, open questions remain about tensions that arise from enacting intersectional moderation, discussed next. 

\section{Discussion}
Moderation is difficult, messy, and emotional. The controversies presented here highlight challenges faced by r/AskHistorians moderators as they engaged in care work to enact an alternative content moderation model on Reddit, a sociotechnical system hostile to their goals \cite{gilbert2020run}. In this section, I use Patricia Hill Collins’ matrix of oppression as a framework to show how r/AskHistorians moderators engaged in care work across levels of power, sometimes reinforcing power, and sometimes resisting it. For example, moderators engaged in care work across personal (within the moderation team), community (r/AskHistorians and its mission of public history), and systemic (the platform and beyond) levels. This section is organized by each layer of power. Within each subsection, I discuss how power manifests in moderation, the care work community moderators use to navigate it, and implications for design and policy. 

\subsection{Intersectional moderation at the personal level}

Collins \cite{collins1990black} notes that while it is often assumed that dominance is asserted from the top down, it also manifests through interpersonal relationships, which can be empowering or oppressive. Relationship development between moderators was an integral part of the r/AskHistorians experience for many in the moderation community. Friendships made it easier to manage external conflicts, such as animosity from users or ambivalence from Reddit's administrators, and provided social support that helped prevent burnout \cite{gilbert2020run, dosono2019moderation, dosono2020decolonizing}. The importance of friendship can be seen in the second controversy, in which the moderation team joked with each other in response to the onslaught of racism and abuse. It's also present in the third controversy, during a highly stressful situation (although, as Nick notes in his reflection, joking may also feel alienating and be elitist). Additionally, although the majority of moderation actions are conducted by individuals working independently, the hardest moderation tasks, like making decisions about borderline content, are often tackled collectively to build consensus. Friendships, more often than not, facilitated consensus-building within the mod community. 

However, the first and third controversies show that these friendships can make moderation work more difficult due to power relationships that developed between members of the moderation team. While decisions regarding borderline answers are collaborative, having expertise on topics resulted in the build up of what Jo Freeman \cite{freeman1972tyranny} describes as ``soft power,'' where a subset of people in an ostensibly structureless group occupy informal positions of power. In their reflections, PH and Aaron note how Becca's soft power, grounded in her expertise, influenced how the decision-making process played out in that discussion. When others pushed for more information about why the answer should be removed, Becca was hurt because she felt her expertise wasn't recognized by the team. Trying to thread the needle between engaging in a discussion about a moderation decision while preserving their friendship took days of work. Even though it was largely unsuccessful, they did it because they cared about Becca, the flaired user, and the community. 

Ultimately, they made a decision that prioritized Becca's expertise over the user's and risked sacrificing the relationship with the user to preserve the relationship with Becca. Similarly, in the third controversy, the team opted to focus on mitigating Andy's concerns rather than discuss broader moderation issues around how to manage reactions to injustice that violate typical civility norms. Over time, disagreements such as these and others resulted in the formation of loose ``factions'' (alluded to by Jim). In these factions, some team members prioritized expertise and relationships between moderators, resulting in a stricter approach to moderation while others prioritized relationships with users, resulting in a more lenient approach to moderation. As in other non-profit organizations \cite{ashforth2014functions} that contend with splits between ``idealism'' (the stricter approach) and ``pragmatism'' (the broader approach), the r/AskHistorians team experienced moments of intense conflicts centering on ``lightning rods'' who were viewed as prototypical of each faction, exemplified in the third controversy through Dave's attack on Claire. 

\subsubsection{Supporting care work at personal levels of power}
The care involved in managing relationships and navigating power structures in the course of maintaining the community increases the work involved in moderation. The discussions described in the first and third controversies spanned days, and the emotional toll on moderators navigating relationship maintenance was considerable. The arduous nature of moderation is reflected in comments made by Lily, Kevin, and I in the third controversy, which all describe conflicts with team members as highly stressful and echo recent findings that interpersonal conflicts are among the top reasons why volunteer moderators quit \cite{schopke2022why}. This care work, and the impact it has on moderators, is largely invisible to those outside the moderation team, making it challenging for designers and policy makers to account for. Further, it can disrupt moderators' ability to account for intersectionality in moderation practices. Because of its important role in supporting challenging moderation work and avoiding burnout, policy makers should support moderation models that allow for collaboration between moderators. However, because this work also risks exacerbating disruption and potentially lead moderators to make decisions that undermine the goals of intersectional moderation, policy makers and designers should support systems and processes, such as mediation from non-team members, that help moderators navigate conflicts. 

\subsection{Intersectional moderation at the community level}

Collins \cite{collins1990black} describes community-based power as formed by experiences and ideas shared with members of a group. This power is exerted by dominant groups, which aim to replace the knowledge held by subjected groups, to simplify control. Common moderation models used by platforms are applied broadly (e.g., through rules and community standards that apply across the entire platform), without accounting for power differentials between groups, which can support the proliferation of hate, harassment, and abuse \cite{zittrain2022how, schoenebeck2020reimagining, phillips2021you}. Over-reliance on common moderation models has led to calls for moderation models that distribute power from dominant groups, including community moderation \cite{zittrain2022how, christian2020platforming}. These models would account for context and nuance, ideally preventing blanket (and potentially incorrect or unfair) decisions \cite{schoenebeck2020reimagining}. However, while community content moderation is more democratic than commercial moderation models, it remains a top-down model of moderation in which a small subset of people control content and determine who can participate. 

Community moderators occupy a contentious role at the community level of power. On the one hand, they work within a system more powerful than themselves. On the other hand, they occupy positions of power relative to users \cite{matias2019civic}. For r/AskHistorians moderators, the role of moderation is further complicated by its mission: public history. The key role of the public history mission is apparent across all three controversies: the importance of maintaining high quality content in the first; the willingness to create high quality content in the second; and fear of the community collapsing in the third. Oliver used a gardening metaphor to highlight his desire to nurture the community through moderation work—a desire expressed by other community moderators \cite{seering2022metaphors} that shows the deep care moderators have for their communities. 

To support r/AskHistorians as a public history site, its moderators focus on building trust with their audiences. They work in a system lacking traditional indicators of expertise, such as occupation or advanced degree, which places heightened importance on the content of answers \cite{gilbert2020run, breit2018how, raeburn2022out}. In turn, this heightens the importance of moderation decisions, because allowing an incorrect answer to remain risks eroding trust among community members, thereby undermining its mission of public history. Some r/AskHistorians moderators developed a high standard for what counts as ``high-quality,'' ``in-depth,'' and ``comprehensive,'' resulting in what the moderation team described as ``standards creep.'' 

Although maintaining quality is vital to r/AskHistorians’ public history mission, so too is motivating participation \cite{kraut2012building}. It takes time to write a response that satisfies requirements for depth and comprehension \cite{gilbert2020run}; further, there are a limited number of people with expertise on a given topic area on Reddit. As a result, many questions are left unanswered. So while high standards help ensure trust, low response rates can erode it, particularly if removals are perceived to be unfair \cite{tyler2021social}. Eroding trust creates competing standards, each of which is tied to the mission—under moderate and breed skepticism, over moderate and smother the community. A tension about the best degree of moderation drove the disagreement in the first controversy, and explains how one decision about one borderline answer could explode into an argument that lasted weeks, and even contribute to interpersonal tensions that spanned years. 

Making good decisions about what content to remove plays an important role in building trust in the knowledge shared within the community. However, it also creates tensions between moderators and users whose comments are removed. Maintaining relationships with users, particularly the community's experts, supports r/AskHistorians' public history mission. As can be seen in the first controversy, moderators routinely provide notices to users when their answers are in-depth or comprehensive, but otherwise violate a rule. While research shows that public communication of norms can decrease rule-violating behavior \cite{tankard2016norm} including harassment and abuse \cite{matias2019preventing}, on r/AskHistorians removals of in-depth conduct are typically communicated privately to users. Private communication protects users. For the question asker, private comment removals prevent the discussion from digressing from the question, and for the commenter, ``rebukes'' are not made public. Providing these removal notices involves significant labor and expertise, as highlighted in the reflections shared by moderators in the first controversy, creating what Baym \cite{baym2015connect} describes as ``relational labor'' in which moderators conduct work with the intent to foster ongoing connections. The first controversy highlights how painful this experience can be for question answerers when moderators get communication wrong. To the flaired user in the first controversy, even though notice had been provided, it felt paternalistic and alienating. Simply providing notification is not enough. Providing effective notification takes time, care, and subject-area expertise. 

Although r/AskHistorians moderators routinely provide notifications when answers are in-depth and/or given by users from the expert flair community, the vast majority of removals are not given notice, despite removal notices being more transparent \cite{myers2018censored} and preferred by users \cite{jhaver2019did, jhaver2019does}. This is because sociotechnical elements of Reddit disincentivize moderators from providing removal notifications. Providing public notifications is the quickest and easiest way to notify a user of the removal. Tools such as Moderator Toolbox (a plugin designed by Reddit user, u/creesch) provide easily accessible removal macros that can be left with the click of a button. Providing private notifications requires an additional step, and means that moderators must navigate away from the comment thread they are reviewing to modmail, creating more work. There are also disincentives for providing public notification. Because threads display the total number of comments made rather than those that remain, providing notifications in public adds to the comment counter. When Reddit users, many of whom hold technolibertarian values of free speech \cite{massanari2017gamergate}, anticipate an active thread but see that most comments have been removed, they become frustrated, which can lead to pushback and abuse \cite{gilbert2020run}. The most expedient way to provide notice is also the way that places moderators at greatest risk. 

\subsubsection{Supporting care work and easing tensions caused by community power} 

While balancing decision-making approaches drives the success of the community, moderators have very little information about how decision-making impacts their communities over time. On r/AskHistorians, a small group of moderators puts considerable effort into assessing the health of the community. Once calculated by hand, a script has since been developed that tracks how many questions get answers and charts that data over time. However, the r/AskHistorians moderation team lacks explanatory information about the results. Other moderation teams without the technical expertise (or time) cannot easily monitor longitudinal impacts of decision-making that could help scaffold successful policy development and enforcement. As pushes for community-based moderation models have been increasing (e.g., \cite{zittrain2022how, christian2020platforming}) and some users are moving from platform based-models to federated systems that rely on volunteer moderation labor, platforms that wish to support successful communities should make data and tools available to moderation teams that support longitudinal analyses and allow them to make informed decisions about how to best nurture their communities. Further, future research on transparency in moderation should explore best practices for engaging with users when their content is removed. This is particularly important for supporting intersectional moderation. Recent work by Thach and colleagues \cite{thach2022visible} has shown that moderation visibility is particularly important for marginalized users; for example, supporting connective labor \cite{pugh2022emotions} helps moderators recognize and respond to users' emotions. Ultimately, considerations should account for both users’ experiences as well as moderators,’ for whom providing notifications takes time and increases risk, and may be members of marginalized communities.

\subsection{Intersectional moderation at the systemic level} 
At the systemic level, power is held by social institutions that expose individuals to thoughts representing the dominant viewpoint \cite{collins1990black}. On Reddit, dominant viewpoints are contributed by its default white male demographic and propagated through its voting system \cite{massanari2017gamergate}. r/AskHistorians uses an intersectional moderation model to subvert that system: first, through content removals that provide space for non-dominant viewpoints and time for people to write in-depth and comprehensive answers to questions; and second, through proactive moderation, such as responding to questions \cite{gilbert2020run} and, as shown in the second controversy, by preemptively providing historical context to current events. 

The second story showcases how an intersectional model can resist systemic power. Because r/AskHistorians is shaped by the questions its predominantly white male userbase asks \cite{gilbert2020run}, organic question-asking processes often center white male histories; while upvoting (also conducted by the largely white male user base) creates patterns of weak-tie racism \cite{brock2020distributed} in which the algorithms that drive Reddit create microagressions at scale. In the weeks after the murder of George Floyd, this pattern was replicated. To counter this pattern and re-center Black history, the moderation team used proactive moderation \cite{habib2022proactive} by collaborating on a response, thus providing a source of information users may have not have otherwise encountered, creating space in which Reddit and r/AskHistorians' default whiteness was interrupted. The second story also illustrates how moderators were able to use their position of power relative to users to remove hate speech, hate-based disinformation, and microaggressions, thus prioritizing positive freedom \cite{phillips2021you}. 

However, any online community’s ability to successfully implement intersectional moderation models is limited by the sociotechnical systems in which they operate. On Reddit, these systems include voting and moderators’ ability to account for context. While the second story highlights successes of proactive content moderation, similar initiatives on r/AskHistorians have not had the same effect due to the voting system. For example, a post co-written by two Indigenous moderators on Columbus Day received upvotes, but one moderator's comments supporting reparations was downvoted. His position obfuscated from view in a form of technological colonialism perpetuated through Reddit's voting system, which prioritizes the feelings and emotions of the dominant group \cite{bonilla2006racism, brock2020distributed}, replicating colonialist patterns identified on other platforms, such as Quora \cite{das2021jol}, and other subreddits \cite{wu2022conversations}. In the third controversy, the original argument, which involved a user accusing a former moderator of being racist, highlights both the importance of accounting for context and moderators’ ability to do so successfully if they wish to subvert rather than replicate systemic power. As Claire noted, supporting equity through moderation requires context; otherwise, as Gray and Stein \cite{gray2021we} have found, moderation risks replicating carceral structures. However, on Reddit, that context is often unavailable. While I disagreed with Viktoria's point about the use of post histories as a potential source of contextual information at the time, she was right. Post histories on Reddit, which primarily contain textual information, can easily be faked in an attempt to perpetuate discrimination. Even when using intersectional moderation that attempts to subvert systemic power on Reddit, moderators contend with features that silence already marginalized voices, and risk silencing marginalized voices themselves through good-intentioned removals. 

\subsubsection{Supporting care work at systemic levels of power}
If intersectional moderation models are going to successfully counter systemic power, policy makers developing these models need to account for the sociotechnical systems through which these models would be enacted. For example, while democratic or distributed moderation models can provide powerful opportunities to involve users in processes that affect them, they risk crowding out minority and marginalized voices and reinforcing systemic power structures. For example, Wu and colleagues \cite{wu2022conversations} found that members of the r/baltimore community who attempted to disrupt racist narratives were often downvoted, while racist comments were upvoted. Careful designing of these systems, such as weighting the impact of votes depending on the source, may help mitigate effects like those experienced by r/AskHistorians’ Indigenous moderators described above. 

While it's tempting to increase the visibility of users to account for systemic power, obfuscation is a vital protective tactic on Reddit. For example, Viktoria obfuscates her identity as a transgender woman to protect her from abuse. Designers who would like to support contextual moderation without compromising users’ safety can consider how selective or flexible visibility may help. For example, r/BlackPeopleTwitter has a system in which Black, Indigenous, and People of Color (BIPOC) can send the moderation team a picture of their arm and verified users are then allowed to participate in ``Country Club Threads.'' While classification based on phenotypical characteristics can be problematic \cite{bowker2000sorting}, it nonetheless provides some support for allowing people to preserve their anonymity while freely participating in a space that centers Black users and experiences while limiting disruptive participation by outsiders and racists \cite{klassen2021more}. In other words, designers should consider systems that allow users to be visible to certain groups. Similarly, in her work with transfeminine TikTok users who rely on visibility, DeVito \cite{devito2022how} describes how transfemmes use folk theorization to navigate visibility traps, creating both actionable theories that allow them to safely be visible and demotivational theories that alert creators to risks. Drawing from these theories, designers could also consider developing systems that enable users to be visible when they want to be, and not when they don’t.  

\section{Conclusion}
Current models of moderation are not working. In response, policy and law makers, scholars, and technologists are seeking alternative models of content moderation to address these issues by rethinking how we approach moderation. For example, how do we address harm \cite{schoenebeck2020reimagining} and is it possible to do so proactively? \cite{habib2022proactive} What happens when we center positive freedoms for the communal good over negative freedoms for the individual in such a way that doesn't replicate harm? \cite{phillips2021you}. Community moderators are confronted with these are questions, and how to successfully enact solutions every day. Learning more about the decisions they make, the challenges they face, and where they are successful (and where they flounder) provides valuable insight that can help anticipate sticking points for other community moderators and platforms applying alternative models. 

If we want to address harm online and create fairer online spaces moderation needs to account for power. By drawing from Black feminist theories of power, and intersectionality in particular \cite{collins1990black, crenshaw2018demarginalizing}, we can learn more about how power manifests across various levels, and how it affects people who are marginalized by moderation, for example, due to race, gender, sexuality, class, and disability. r/AskHistorians moderators use an alternative moderation model to support public history and in doing so regularly negotiate power—at times reinforcing it and at others resisting it. The controversies presented in this paper show how much work and care goes into navigating power and how interpersonal relationships, platform design, and policies can stymie moderators' ability to account for power through their moderation practices. To advance an intersectional model of moderation through which moderators can support resistance and resilience more work needs to be done. Key to this will be collaborations with moderators and the moderated.

\begin{acks}
I have a lot of thanks to give. First, to the incredible reviewers, who provided detailed, encouraging, and thoughtful reviews. Second, to Aure Schrock of Indelible Voice for their monumental editorial services. But most of all to the moderators of r/AskHistorians. Thank you for trusting me and welcoming me to your community as a researcher, for your ongoing support of this project, and for your friendship. 

This project was supported by funding from NSF \#2131508 and TWCF-2020-20411. 
\end{acks}

\bibliographystyle{ACM-Reference-Format}
\bibliography{AskHistorians}


\appendix
\section{Additional Methodological Notes}

\subsection{Voting}
As noted above, I requested a vote be held to determine whether or not the controversies should be published. I held the vote before comments were added so that no one would spend time writing annotations on a paper the team agreed not to publish. Of the 31 active moderators eligible to vote, 21 voted for publication, 2 (including me) abstained from the vote, and 0 voted against publication. When a full draft of the paper was complete, I shared it with the mod team again for further review, although a second vote was not held. 

\subsection{Entree and the Covid-19 pandemic}
I began data collection the last week of January 2020. Just over a month later, the world shut down due to the Covid-19 pandemic. Initially, I had attempted to maintain a level of distance between myself and the moderation team. Fear and uncertainty related to the pandemic rose and lockdowns increased in-person isolation, leading many members of the moderation team to share their anxieties, fears, and experiences with the others. Within the first few weeks of the shutdown, one moderator lost her grandfather and later caught Covid herself. She was the first person I knew who had it. Another was laid off. One spent weeks in terrible pain—he had a tooth infection but couldn’t get an appointment with a dentist. I lost my grandfather too, but I didn’t tell anyone. For a week I couldn't stop sobbing whenever I got in the shower. But as moderators on the team opened up in the wake of the pandemic, I began to struggle with how much I should be sharing. Going into the project I had only considered extraction in terms of labor—was it right or ethical for me to bear witness to emotional experiences without also making myself vulnerable? So, I started sharing. For example, when we lost our apartment in the middle of lockdown I told the team. One of the mods helped us find a new place to live. At some point (I’m not sure when) I noticed my language began to shift—when I talked about r/AskHistorians I began saying ``we'' instead of ``they.'' At some point (I’m not sure when) I was no longer just an ethnographer studying moderation, I was an r/AskHistorians moderator. I am part of the stories shared above. 

\end{document}